\documentclass[aps,prl,%
twocolumn,%
amssymb,amsmath,%
showpacs,showkeys,%
tightenings,%
ejectum,%
breveted]{revtex4}
\usepackage[final]{graphicx}
\usepackage{dcolumn}
\begin{document}
\title{Systematically Accelerated Convergence of Path Integrals}
\author{A. Bogojevi\'c}
\author{A. Bala\v{z}}
\author{A. Beli\'c}
\affiliation{Institute of Physics, P.O. Box 57, 11001
Belgrade, Serbia and Montenegro}
\begin{abstract}
We present a new analytical method that systematically improves the convergence of path integrals of a generic $N$-fold discretized theory. Using it we calculate the effective actions $S^{(p)}$ for $p\le 9$ which lead to the same continuum amplitudes as the starting action, but that converge to that continuum limit as $1/N^p$. We checked this derived speedup in convergence by performing Monte Carlo simulations on several different models. 
\end{abstract}
\preprint{SCL preprint}
\maketitle


Since their inception \cite{feynmanhibbs,feynman} path integrals have presented an extremely compact and rich formalism for dealing with quantum theories. They have grown into powerful tools for dealing with symmetries (including gauge symmetry), for deriving non-perturbative results (e.g. solitons, instantons, symmetry breaking) and for showing connections between different theories or different sectors of the same theory (e.g. bosonization, duality) \cite{itzyksonzuber,coleman}. They have also consistently allowed us to extend and generalize quantization procedures to ever more complicated systems. Path integrals have brought about a rich cross fertilization of ideas between high energy and condensed matter physics by delineating and strengthening similarities that exist between statistical and quantum systems \cite{itzyksondrouffe,parisi}. Today, path integrals are used both analytically and numerically \cite{barkerhenderson,barker,kaloswhitlock,pollockceperley,ceperley} in many other areas of physics, chemistry and materials science. They are starting to play more prominent roles in several areas of mathematics and in modern finance \cite{baaquie}. An extensive review of path integrals and their applications can be found in \cite{kleinert}. 

Unfortunately, we still have very little knowledge of the mathematical properties of path integrals. In addition, a very small number of path integrals can be solved exactly. Although the functional formalism has been used to derive many general approximation techniques (e.g. perturbation, semi-classical expansion, variational methods) along with many model-specific approximation techniques, there still remains a wealth of interesting models that can't be analyzed analytically and need to be treated numerically. The definition of path integrals as a limit of multiple integrals makes their numerical evaluation quite natural. The most all-around applicable numerical method for such calculations is based on Monte Carlo simulations. However, numerical integration of path integrals is notoriously demanding of computing time -- so much so that specific path integral calculations serve as benchmarks for new generations of supercomputers. 

Several research groups have focused on improving the convergence of path integrals. The best available result for a generic theory (valid only for partition functions and not for general amplitudes) is the convergence of $N$-fold discretized expressions as $1/N^4$ \cite{takahashiimada,libroughton,jangetal}. Related investigations have focused either on improvements in short time propagation \cite{makrimiller1,makrimiller2,makri} or have presented model specific improvements of the action \cite{alfordetal,bondetal}. Ref. \cite{krajewskimuser} gives a useful comparison of several different approaches.

In order to further significantly speed up numerical procedures for calculating path integrals for a generic theory it is necessary to add new analytical input. In this letter we present a systematic investigation of the relation between different discretizations of a given theory. A result of this investigation is a procedure for constructing a series of effective actions $S^{(p)}$ having the same continuum limit as the starting action $S$, but which approach that limit as $1/N^p$. We use this procedure to obtain explicit expressions for these effective actions for $p\le 9$. 

In the functional formalism the quantum mechanical amplitude $A(a,b;T)=\langle b|e^{-T\hat H}|a\rangle$ is given 
in terms of a path integral which is simply the $N\to\infty$ limit of the $(N-1)$-- fold integral expression
\begin{equation}
\label{amplitudeN}
A_N(a,b;T)=\left(\frac{1}{2\pi\epsilon_N}\right)^{\frac{N}{2}}\int dq_1\cdots dq_{N-1}\,e^{-S_N}\ .
\end{equation}
The euclidean time interval $[0,T]$ has been subdivided into $N$ equal time steps of length $\epsilon_N=T/N$, with $q_0=a$ and $q_N=b$. $S_N$ is the naively discretized action of the theory. We focus on actions of the form 
\begin{equation}
\label{action}
S=\int_0^Tdt\,\left(\frac{1}{2}\, \dot q^2+V(q)\right)\ ,
\end{equation}
whose naive discretization is simply
\begin{equation}
\label{actionN}
S_N=\sum_{n=0}^{N-1}\left(\frac{\delta_n^2}{2\epsilon_N}+\epsilon_NV(\bar q_n)\right)\ ,
\end{equation}
where $\delta_n=q_{n+1}-q_n$, and $\bar q_n=\frac{1}{2}(q_{n+1}+q_n)$. We use units in which $\hbar$ and particle mass equal 1.

The definition of the functional integral makes it necessary to make the transition from the continuum to the discretized theory, a process that is far from unique. For theories described by eq.~(\ref{action}) we have the freedom to choose any point in $[q_n,q_{n+1}]$ in which to evaluate the potential without changing physics -- the discretized amplitudes do differ, but they tend to the same continuum limit. The calculations we present turn out to be simplest in the mid-point prescription where $V$ is evaluated at $\bar q_n$. A more important freedom related to our choice of discretized action has to do with the possibility of introducing additional terms that explicitly vanish in the continuum limit. Actions with such additional terms will be called effective. For example, the term 
$\sum_{n=0}^{N-1}\epsilon_N\,\delta_n^2\,g(\bar q_n)$, 
where $g$ is regular when $\epsilon_N\to0$, does not change the continuum physics since it goes over into $\epsilon_N^2\int^T_0dt\,\dot q^2\,g(q)$, i.e. it vanishes as $\epsilon_N^2$. Such terms do not change the physics, but they do affect the speed of convergence. A systematic study of the relation between different discretizations of the same path integral will allow us to explicitly construct a series of effective actions with progressively faster convergence to the continuum. 

We start by studying the relation between $2N$-fold and $N$-fold discretizations. From eq.~(\ref{amplitudeN}) we see that we can write the $2N$-fold amplitude as an $N$-fold amplitude given in terms of a new action $\widetilde S_N$ determined by
\begin{equation}
\label{tildeS}
e^{-\widetilde S_N}=\left(\frac{2}{\pi\epsilon_N}\right)^{\frac{N}{2}}\int dx_1\cdots dx_N\,e^{- S_{2N}}\ ,
\end{equation}
where $S_{2N}$ is the $2N$-fold discretization of the starting action. We have written the $2N$-fold discretized coordinates $Q_0,Q_1,\ldots,Q_{2N}$ in terms of $q$'s and $x$'s in the following way: $Q_{2k}=q_k$ and $Q_{2k-1}=x_k$. Note that we have $q_0=a$, $q_N=b$, while the $N-1$ remaining $q$'s play the role of the dynamical coordinates in the $N$-fold discretized theory. The $x$'s are the $N$ remaining intermediate points that we integrate over in eq.~(\ref{tildeS}). 

In order to iterate this discretization halving process the new action $\widetilde S_N$ must belong to the same class of expressions as the starting actions $S_N$. It is not difficult to show that the naively discretized action does not satisfy this requirement, i.e. integration of eq.~(\ref{tildeS}) yields new types of terms in $\widetilde S_N$. In fact, the class of actions closed to transformation (\ref{tildeS}) is of the form
\begin{eqnarray}
\label{actions}
\lefteqn{
S_N=\sum_{n=0}^{N-1}\bigg(\frac{\delta_n^2}{2\epsilon_N}+\epsilon_N\,V(\bar q_n)+\epsilon_N\,\delta_n^2\,g_1(\bar q_n)\,+}\nonumber\\
&&{}\quad+\,\epsilon_N\,\delta_n^4\,g_2(\bar q_n)+\epsilon_N\,\delta_n^6\,g_3(\bar q_n)+\ldots\bigg)\ .
\end{eqnarray}
All the functions appearing above also depend on the time step $\epsilon_N$. We will not display this dependence explicitly in order to have a more compact notation. All of the functions are regular in the $\epsilon_N\to 0$ limit making these effective actions equivalent to the starting action. From eq.~(\ref{tildeS}) and (\ref{actions}) we can derive the following integral relation which determines the new action $\widetilde S_N$ in terms of the starting action:
\begin{widetext}
\begin{equation}
\label{integral}
\exp\bigg(-\epsilon_N\Big(\widetilde V(\bar q_n)+\delta_n^2\,\widetilde g_1(\bar q_n)+\delta_n^4\,\widetilde g_2(\bar q_n)+\ldots\Big)\bigg)=
\left(\frac{2}{\pi\epsilon_N}\right)^{\frac{1}{2}}\int_{-\infty}^{+\infty} 
dy\,\exp\left(-\frac{2}{\epsilon_N}y^2\right)F(\bar q_n+y)\ ,
\end{equation}
\begin{equation}
\label{F}
-\,\frac{\ln F(x)}{\epsilon_N}=\frac{1}{2}\,V\Big(\frac{q_{n+1}+x}{2}\Big)+\frac{1}{2}\,V\Big(\frac{x+q_n}{2}\Big)+
\frac{(q_{n+1}-x)^2}{2}\,g_1\Big(\frac{q_{n+1}+x}{2}\Big)+\frac{(x-q_n)^2}{2}\,g_1\Big(\frac{x+q_n}{2}\Big)+\ldots\ .
\end{equation}
\end{widetext}

The above integral equation can be solved for the simple cases of a free particle and a harmonic oscillator, and gives the well known results. 
Note however that for a general case the integral in eq.~(\ref{integral}) is in a form that is ideal for an asymptotic expansion \cite{erdelyi}. The time step $\epsilon_N$ is playing the role of small parameter (in complete parallel to the role $\hbar$ plays in standard semi-classical, or loop, expansion). As is usual, the above asymptotic expansion is carried through by first Taylor expanding $F(\bar q + y)$ around $\bar q$ and then by doing the remaining Gaussian integrals. Assuming that $\epsilon_N<1$ (i.e. $N>T$) we have
\begin{eqnarray}
\label{formula}
\lefteqn{\widetilde V(\bar q_n)+\delta_n^2\,\widetilde g_1(\bar q_n)+\delta_n^4\,\widetilde g_2(\bar q_n)+\ldots=}\nonumber\\
&&=-\frac{1}{\epsilon_N}\ln\left[\sum_{m=0}^{\infty}\frac{F^{(2m)}(\bar q_n)}{(2m)!!}\,\left(\frac{\epsilon_N}{4}\right)^m\right]\ .
\end{eqnarray}
All that remains is to calculate the $F^{(2m)}(\bar q_n)$'s using eq.~(\ref{F}) and to expand the potential and all the functions $g_k$ 
around the mid-point $\bar q_n$. This is a straight forward though tedious calculation. In this letter we illustrate the general procedure for calculating $\widetilde S$ by explicitly giving its expansion to order $\epsilon_N^3$:
\begin{eqnarray}
\label{halvingrelations}
\widetilde V &=&
V+\epsilon_N\left[{\frac{1}{4}}g_1+\frac{1}{32}V''\right]+\nonumber\\
&&+\epsilon_N^2\,\left[\frac{3}{16}g_2-\frac{1}{32}V'\,^2+\frac{1}{2048}V^{(4)}+\frac{3}{128}g_1''\right]\nonumber\\
\widetilde g_1 &=& \frac{1}{4}g_1+\frac{1}{32}V''+\\
&&+\epsilon_N\left[\frac{3}{8}g_2+\frac{1}{1024}V^{(4)}-\frac{1}{64}g_1''\right]\nonumber\\
\widetilde g_2 &=& \frac{1}{16}g_2+\frac{1}{6144}V^{(4)}+\frac{1}{128}g_1''\ .\nonumber
\end{eqnarray}

In the above relations we expanded $\widetilde V$ up to $\epsilon_N^2$, $\widetilde g_1$ up to $\epsilon_N$, etc. We also disregarded all the higher $\widetilde g_k$'s. The reason for this is that the short time propagation of any theory satisfies $\delta_n^2\propto\epsilon_N$ while the $g_k$ term enters the action multiplied by $\delta_n^{2k}$. In general, if we expand the new action $\widetilde S$ to $\epsilon_N^p$ we need to evaluate only $\widetilde V$ (up to $\epsilon_N^{p-1}$) and the first $p-1$ functions $\widetilde g_k$ (up to $\epsilon_N^{p-1-k}$). The task of calculating $\widetilde S$ to large powers of $\epsilon_N$ is time-consuming and is best done with the help of a standard package for algebraic calculations such as Mathematica. Using Mathematica we determined the corresponding expressions for $p\le 9$. The above solution for $\widetilde S$ (and its analogues for higher values of $p$) has the following important property: up to $O(\epsilon_N^p)$, a coarser $N$-fold discretization using $\widetilde S$ does the same job as the $2N$-fold discretization of the starting theory, i.e. $\widetilde A_N(a,b;T) = A_{2N}(a,b;T)+O(\epsilon_N^p)$.

We next iterate the discretization halving process outlined above in order to connect up the $2^s N$-fold and $N$-fold discretizations. Ultimately we will focus on the solution when $s\to\infty$, i.e. the one connecting the continuum theory with its $N$-fold discretization. For the $p=3$ case that we continue to use as an illustration of the general procedure, the above iterative process is governed by:
\begin{widetext}
\begin{eqnarray}
\label{rec:p3}
V_{k+1} &=&
V_{k}+\frac{\epsilon_N}{2^{s-k-1}}\left[{\frac{1}{4}}(g_1)_{k}+\frac{1}{32}V_{k}''\right]+
\frac{\epsilon_N^2}{2^{2(s-k-1)}}\,\left[\frac{3}{16}(g_2)_{k}-\frac{1}{32}V_{k}'\,^2+\frac{1}{2048}V_{k}^{(4)}+\frac{3}{128}(g_1)_{k}''\right]\nonumber\\
(g_1)_{k+1} &=& \frac{1}{4}(g_1)_{k}+\frac{1}{32}V_{k}''+
\frac{\epsilon_N}{2^{s-k-1}}\left[\frac{3}{8}(g_2)_{k}+\frac{1}{1024}V_{k}^{(4)}-\frac{1}{64}(g_1)_{k}''\right]\\
(g_2)_{k+1} &=& \frac{1}{16}(g_2)_{k}+\frac{1}{6144}V_{k}^{(4)}+\frac{1}{128}(g_1)_{k}''\ ,\nonumber
\end{eqnarray}
\end{widetext}
where $k=0,1,2,\ldots,s-1$. The zeroth iterate corresponds to the starting action, the last iterate to the effective action whose $N$-fold discretization is equivalent to the $2^s N$-fold discretization of the starting theory. The $\epsilon_N/2^{s-k-1}$ terms represent the time step of the $k$-th iterate in the discretization halving procedure. 

Although the above system of recursive relations is non-linear, it is in fact quite easy to solve if we remember that the system itself was derived via an expansion in $\epsilon_N$. Having this in mind we first write all the functions as expansions in powers of $\epsilon_N$ that are appropriate to the level $p$ we are working at. For $p=3$, we have   
\begin{eqnarray}
\label{expand}
V_k &=& A_k + \frac{\epsilon_N}{2^{s-k-1}}B_k+\left(\frac{\epsilon_N}{2^{s-k-1}}\right)^2C_k\nonumber\\
(g_1)_k &=& D_k + \frac{\epsilon_N}{2^{s-k-1}}E_k\\
(g_2)_k &=& F_k\ .\nonumber
\end{eqnarray}
Putting this into eq.~(\ref{rec:p3}) we find that $A_{k+1}=A_k$. Since $A_0=V$ it follows that $A_k=V$ for all $k$. The remaining terms obey a linear system of equations that is easily solved for given initial conditions. Rather than solving the system and then taking the $s\to\infty$ limit of the solutions, it is easier to  set $k=s-1$ in the system and then take the limit $s\to\infty$. By doing this we directly obtain the continuum limit solution of the discretization halving process connecting the continuum theory to its $N$-fold discretization. The continuum limit solution of the $p=3$ level system (the solution of the continuum limit of the recursive relations given in eq.~(\ref{rec:p3}) describing discretization halving at the $p=3$ level) is:
\begin{eqnarray}
\label{p3continuum}
V_{p=3} &=& V + \epsilon_N\frac{V''}{12} + \epsilon_N^2\left[-\frac{V'\,^2}{24}+\frac{V^{(4)}}{240}\right]\nonumber\\
(g_1)_{p=3} &=& \frac{V''}{24} + \epsilon_N\frac{V^{(4)}}{480}\\
(g_2)_{p=3} &=& \frac{V^{(4)}}{1920}\ .\nonumber
\end{eqnarray}
Note that the continuum solution depends only on the initial potential $V$, i.e. it is not sensitive to initial values of the $g_k$'s as these terms all vanish in the continuum limit. In this way we have obtained the effective action that gives the best $N$-fold discretization of the starting theory at the $p=3$ level -- the one differing from the continuum limit by a term of order $1/N^3$. One can similarly obtain continuum limit effective actions $S^{(p)}$ for higher $p$. These effective actions lead to $N$-fold amplitudes that deviate from the continuum amplitude as $O(\epsilon_N^p)$ 
\begin{equation}
\label{convergence}
A^{(p)}_N(a,b;T) = A(a,b;T)+O(\epsilon_N^p)\ .
\end{equation}

It is important to note that one solves for the continuum limit of the level $p$ system of recursive relations but once for all theories, i.e. once the solution is found it works for all potentials $V$. The only requirement for the level $p$ solution is that the starting potential is differentiable $2p-2$ times. Solutions for larger values of $p$ are a bit more cumbersome, however, they are just as easy to use in simulations. We have found that the growth in complexity of the effective actions with increasing $p$ has little effect on computation time for $p\le 4$, while simulations with $p=9$ are roughly ten times slower due to this. However, this is an extremely small price to pay for a gain of nine orders of magnitude in the speed of convergence. Expressions for effective actions up to $p=9$ can be found on our web site \cite{scl}. 
\begin{figure}[!ht]
    \includegraphics[width=8.cm]{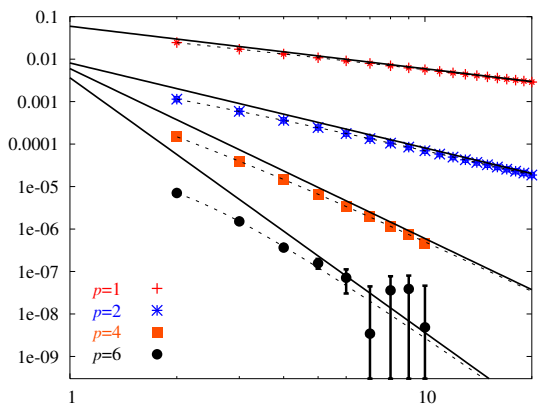}
    \caption{\label{p1246} (Color online) Deviations from the continuum limit $|A_N^{(p)}-A|$ as functions of $N$ for $p=1,2,4$ and $6$ for an anharmonic oscillator with quartic coupling $\lambda=10$, time of propagation $T=1$ from $a=0$ to $b=1$. $N_{MC}$ was $9.2\cdot 10^9$ for $p=1,2$, $9.2\cdot 10^{10}$ for $p=4$, and $3.68\cdot 10^{11}$ for $p=6$. Dashed lines correspond to appropriate $1/N$ polynomial fits to the data. Solid lines give the leading $1/N$ behavior. The level $p$ curve has a $1/N^p$ leading behavior.}
 \end{figure}

The analytical derivations presented work equally well in both the Euclidean and Minkowski formalism (with appropriate $i\epsilon$ regularization), i.e. they are directly applicable to quantum systems as well as to statistical ones. However, the Monte Carlo simulations used to numerically document our analytical results necessarily needed to be done in the Euclidean formalism. We analyzed in detail several models: the anharmonic oscillator with quartic coupling $V=\frac{1}{2}\,q^2+\frac{\lambda}{4!}\, q^4$ and a particle moving in a modified P\"oschl-Teller potential over a wide range of parameters. In all cases we found agreement with eq.~(\ref{convergence}). Fig.~\ref{p1246} illustrates this behavior in the case of an anharmonic oscillator. We see that the $p$ level data indeed differs from the continuum amplitudes as a polynomial starting with $1/N^p$. The deviations from the continuum limit $|A_N^{(p)}-A|$ become exceedingly small for larger values of $p$ making it necessary to use ever larger values of $N_{MC}$ so that the MC statistical error does not mask these extremely small deviations. For $p=6$ we see that although we used an extremely large number of MC samples ($N_{MC}=3.68\cdot 10^{11}$) the statistical errors become of the same order as the deviations already at $N\gtrsim 8$. For $p=9$ this is the case even for $N=2$, i.e. we already get the continuum limit within a MC error of around $10^{-8}$. 

To conclude, we have presented an algorithm that leads to significant speedup of numerical procedures for calculating path integrals. The increase in speed results from new analytical input that comes from a systematic investigation of the relation between discretizations of different coarseness. We have presented an explicit procedure for obtaining a set of effective actions $S^{(p)}$ that have the same continuum limit as the starting action $S$, but which approach that limit ever faster. Amplitudes calculated using the $N$-point discretized effective action $S_N^{(p)}$ satisfy $A^{(p)}_N(a,b;T)=A(a,b;T)+O(1/N^p)$. We have obtained and analyzed the effective actions for $p\le 9$ and have documented the speedup up to $1/N^9$ by conducting Monte Carlo simulations of several different models. Extension to $d>1$ is in progress. The derivation of the analogue of integral eq.~(\ref{integral}) does not seem to present a problem. The asymptotic expansion used to solve it is also directly generalizable. However, the algebraic recursive relations that determine $S^{(p)}$ will be more complex and may practically limit us to smaller values of $p$. 

We wish to acknowledge financial support from the Ministry of Science and Environmental Protection of the Republic of Serbia through projects 1486 and 1899.

\end{document}